\documentstyle[12pt]{article}
\newcommand{\beq}{\begin{equation}}             
\newcommand{\eeq}{\end{equation}}               
\newcommand{\bqry}{\begin{eqnarray}}            
\newcommand{\eqry}{\end{eqnarray}}              
\newcommand{\bqryn}{\begin{eqnarray*}}          
\newcommand{\eqryn}{\end{eqnarray*}}            
\newcommand{\preprint}[1]{\begin{table}[t]      
            \begin{flushright}                  
            \begin{large}{#1}\end{large}        
            \end{flushright}                    
            \end{table}}                        
\newcommand{\PD}[2]                             
    {\frac{\partial^{#2}}{\partial #1^{#2}}}    
\begin{document}
\preprint{LA-UR-98-1843}
\title{Hadron Mass Scaling in Regge Phenomenology\thanks{Presented at the First
International Conference on Parametrized Relativistic Quantum Theory, PRQT '98,
Houston, Texas, Feb 9-11, 1998}}
\author{\\ L. Burakovsky\thanks{E-mail: BURAKOV@QMC.LANL.GOV} \
\\  \\  Theoretical Division, T-8 \\  Los Alamos National Laboratory \\ Los
Alamos NM 87545, USA}
\date{ }
\maketitle
\begin{abstract}
We show that Regge phenomenology is consistent with the only universal scaling
law for hadron masses, $M^\ast /M=(\alpha ^{'}/\alpha ^{'\ast })^{1/2},$ where 
asterisk indicates a finite-temperature quantity. Phenomenological models 
further suggest the following expression of the above scaling in terms 
of the temperature-dependent gluon condensate: $M^\ast /M=(\alpha ^{'}/\alpha 
^{'\ast })^{1/2}=(\langle G_{\mu \nu }^a\rangle ^\ast /\langle G^{a\mu \nu }
\rangle )^{1/4}.$ 
\end{abstract}


\section*{
}
It is known that the properties of hadrons undergo modifications at finite 
temperature or in nuclear medium \cite{med}. Recently considerable amount of 
work has been devoted to the question of the reduction of the vector meson 
masses at finite temperature and/or density from both theoretical and 
experimental viewpoints. It has been suggested that the quenching of the 
longitudinal response in the quasi-elastic electron-nucleus scattering 
\cite{eN} and the enhancement of the $K^{+}N$ scattering cross-section 
\cite{KN} can be understood as possible consequences of the reduction of the 
vector meson mass in nuclear medium. Recent experiments in heavy ion 
collisions which measure the dilepton mass spectrum \cite{ee} seem to support 
this possibility \cite{RHIC}. A detailed theoretical calculation of the 
dilepton mass spectrum which incorporates various dynamical aspects, including
the reduction of the vector meson masses in medium, was done in \cite{LKB}. 

Also, the question of a universal scaling behavior for all hadron masses at 
finite temperature and/or density has been drawn much attention in the 
literature. Using the effective classical Lagrangian which incorporates
approximate scale and chiral invariance \cite{KEO}, and assuming that the 
ground state properties are dominated by the dilatation field which connects 
the chiral symmatry breaking scale with the gluon condensate via the QCD trace
anomaly, Brown and Rho \cite{BR} have suggested the following universal 
scaling,
\beq
\frac{M^\ast (\rho )}{M(\rho )}\simeq \frac{M^\ast (\sigma )}{M(\sigma )}\simeq
\frac{M^\ast (N)}{M(N)}\simeq \frac{f^\ast _\pi }{f_\pi }\simeq \left( \frac{
\langle \bar{q}q\rangle ^\ast }{\langle \bar{q}q\rangle }\right) ^{1/3}\simeq
\left( \frac{\langle G_{\mu \nu }^aG^{a\mu \nu }\rangle ^\ast }{\langle 
G_{\mu \nu }^aG^{a\mu \nu }\rangle }\right) ^{1/4},
\eeq
where $\langle \bar{q}q\rangle $ and $\langle G^a_{\mu \nu }G^{a\mu \nu }
\rangle $ are the quark and gluon condensates, respectively, $f_\pi $ is the 
pion decay constant, and asterisk stands for a temperature- and/or 
density-dependent quantity. Physical implications of this scaling have been
studied for nuclear physics \cite{nucph}, the equation of state for neutron
stars \cite{nstar}, and the equation of state for hot hadronic matter 
\cite{hadrm}. It was, however, argued in \cite{KW} that the universal scaling
behavior (1) is not realized in the Nambu--Jona-Lasinio model with the 
inclusion of a dilatation field for reasonable parameter ranges. 

The purpose of the present paper is to consider the question of the universal
scaling behavior for all hadronic masses from the viewpoint of Regge 
phenomenology.

It is well known that the hadrons composed of light $(n(=u,d),s)$ quarks 
populate linear Regge trajectories; i.e., the square of the mass of a state 
with orbital momentum $\ell $ is proportional to $\ell:$ $M^2(\ell )=\ell /
\alpha ^{'}\;+$ const, where the slope $\alpha ^{'}$ only very weekly depends 
on the flavor content of the states lying on the corresponding trajectory: 
$\alpha ^{'}_{n\bar{n}}=0.88$ GeV$^{-2},$ $\alpha ^{'}_{s\bar{n}}=0.85$ GeV$^{
-2},$ $\alpha ^{'}_{s\bar{s}}=0.81$ GeV$^{-2};$ it therefore may be taken as a
universal slope in the light quark sector, $\alpha ^{'}\approx 0.85$ GeV$^{
-2}.$ In this respect, the hadron masses as populating collinear trajectories 
do exhibit a universal behavior. Hence, it is quite natural to address the 
question of the scaling of the hadron masses in the framework of Regge 
phenomenology. 

Let us therefore consider the case of finite temperature (which does not differ
technically from the case of finite chemical potential), and assume a universal
scaling behavior for all hadron masses which populate collinear trajectories 
(at zero $T):$
\beq
\frac{M^\ast }{M}=\gamma =\gamma (T),
\eeq
where $\gamma $ is the same function of temperature for every hadronic state.
It then follows that the pattern of initial collinear trajectories constrain 
the form of $\gamma $ in Eq. (2). 

Consider three states with $\ell =1,3,5$ which belong to a common trajectory 
at finite $T:$
\beq
1=\alpha ^{'}M^2(\ell =1)+\alpha (0),\;\;\;3=\alpha ^{'}M^2(\ell =3)+\alpha 
(0),\;\;\;5=\alpha ^{'}M^2(\ell =5)+\alpha (0).
\eeq
Therefore,
\beq
M^2(\ell =5)-M^2(\ell =3)=M^2(\ell =3)-M^2(\ell =1)=\frac{2}{\alpha ^{'}}.
\eeq
At finite $T,$ it follows from (2),(4) that
$$M^{\ast 2}(\ell =5)-M^{\ast 2}(\ell =3)=M^{\ast 2}(\ell =3)-M^{\ast 2}(\ell 
=1)$$
\beq
=\gamma ^2\left( M^2(\ell =5)-M^2(\ell =3)\right) =\gamma ^2\left( M^2(
\ell =3)-M^2(\ell =1)\right) =\frac{2\gamma ^2}{\alpha ^{'}}.
\eeq
Since similar considerations are applied to three states of every trajectory, 
in view of (2), one easily concludes that at finite $T$ hadrons still populate
linear Regge trajectories, albeit with different slope: (since spin does not
change with temperature)
$$1=\alpha ^{'\ast }M^{\ast 2}(\ell =1)+\alpha ^\ast (0),\;\;\;3=\alpha ^{
'\ast }M^{\ast 2}(\ell =3)+\alpha ^\ast (0),\;\;\;5=\alpha ^{'\ast }M^{
\ast 2}(\ell =5)+\alpha ^\ast (0),$$
\beq
\alpha ^{'\ast }=\frac{\alpha ^{'}}{\gamma ^2}.
\eeq
Thus, the slopes of collinear hadron trajectories must satisfy a scaling law,
as well as the hadron masses, and
\beq
\frac{M^\ast }{M}=\left( \frac{\alpha ^{'}}{\alpha ^{'\ast }}\right) ^{1/2}.
\eeq
It is now easy to show that trajectory intercepts do not change with $T:$ 
as follows from $\ell =\alpha ^{'}M^2(\ell )+\alpha (0)=\alpha ^{'\ast }M^{
\ast 2}(\ell )+\alpha ^\ast (0)$ and Eq. (7),
\beq
\alpha ^\ast (0)=\alpha (0).
\eeq

Hence, a universal scaling law (2) for all hadron states which populate 
collinear trajectories implies a similar scaling law for the slopes of these
trajectories, Eq. (6), and the constancy of their intercepts.

Accordingly, the question about the explicit form of this scaling law in terms 
of temperature-dependent quark and/or gluon condensates (e.g., Eq. (1)) is 
directly related to the corresponding dependence of the Regge slope on these
condensates.

Although an exact answer to this question is not known to us yet, a preliminary
consideration of this point can be given.

In the MIT bag model, the Regge slope can be related to the bag constant $B$
\cite{JT}:
\beq
\alpha ^{'}=\frac{1}{16\pi ^{3/2}}\left( \frac{3}{2}\right) ^{1/2}\frac{1}{
\sqrt{\alpha _s}\sqrt{B}},
\eeq
which for the phenomenological values $\alpha _s=0.55$ and $B^{1/4}=0.146$ MeV
\cite{JT} gives
\beq
\alpha ^{'}\approx 0.87\;{\rm GeV}^{-2},
\eeq
in good agreement with data. The bag pressure can be interpreted as the 
difference between the energy density of the physical (nonpreturbative) vacuum
outside the bag and perturbative vacuum inside the bag (it is easily seen from
the bag equation of state $p=p_{free}-B,\;\rho =\rho _{free}+B,$ and therefore
$B=1/4\;(\rho -3p)=1/4\;{\rm Tr}\;\!T_{\mu \nu }).$ Since the energy density 
of the physical vacuum is related to the gluon condensate, via the trace 
anomaly, it is clear that $B\propto \langle G_{\mu \nu }^aG^{a\mu \nu }
\rangle ,$ and therefore, in view of (9), $1/\alpha ^{'2}\propto \langle G_{
\mu \nu }^aG^{a\mu \nu }\rangle .$

Also, Nambu \cite{Nam} derived a string-like equation for the path-ordered 
phase factor $U[\sigma ],$
\beq
\left( \frac{\delta }{\delta \sigma _{n\tau }}\frac{\delta }{\delta \sigma ^{
n\tau }}+C\right) U[\sigma ]=0,\;\;\;U[\sigma ]=P\;\!\exp \left( i\int _\sigma
A_\mu dz^\mu \right),\;\;\;A_\mu =\frac{1}{2}g\Sigma A_\mu ^a\lambda _a,
\eeq
with
\beq
C=G_{n\tau }G^{n\tau }
\eeq
is the gluon field tensor in the normal, $n,$ and tangential, $\tau ,$
directions along the string. Identifying Eq. (11) with the string equations 
from the Nambu-Goto action leads further to
\beq
C=-\left( \frac{1}{2\pi \alpha ^{'}}\right )^2.
\eeq
Hence, two above examples suggest that
\beq
\frac{1}{\alpha ^{'2}}\propto \langle G_{\mu \nu }^aG^{a\mu \nu }\rangle .
\eeq
Although an explicit relation of $\alpha ^{'}$ to the condensates is not known
to us yet, Eq. (14) may be considered as a first approximation to the actual
relation. In view of (14), Eq. (7) may be rewritten as
\beq
\frac{M^\ast }{M}=\left( \frac{\alpha ^{'}}{\alpha ^{'\ast }}\right) ^{1/2}=
\left( \frac{\langle G_{\mu \nu }^aG^{a\mu \nu }\rangle ^\ast }{\langle G_{\mu
\nu }^aG^{a\mu \nu }\rangle }\right) ^{1/4}.
\eeq

Following arguments of ref. \cite{Pas}, one can show that the scaling law (7) 
can be generalized to the pion decay constant, via the relation \cite{Pas}
$\alpha ^{'}f_\pi ^2=8\pi \beta \approx $ const, where $\beta $ is a coupling 
constant in front of the $\pi ^{+}\pi ^{-}\rightarrow \pi ^{+}\pi ^{-}$ 
Veneziano scattering amplitude. An explicit expression of this scaling law in 
terms of the quark condensate (or both the gluon and quark condensates) is 
still an open question. 
  
Finally, we remark that, in view of the scaling law (7), it seems quite 
natural to express the hadron masses in terms of the Regge slope $\alpha ^{'}.$
Such relations for the hadron masses were derived in \cite{prqt} for mesons and
baryons (below we present relations for the ground states only),
\beq
M^2(\rho )=\frac{1}{2\alpha ^{'}},\;\;\;M^2(K^\ast )=\frac{11}{16\alpha ^{'}},
\;\;\;M^2(\phi )=\frac{7}{8\alpha ^{'}},
\eeq
\beq
M^2(N)=\frac{3}{4\alpha ^{'}},\;\;\;M^2(\Sigma ^{'})=\frac{9}{8\alpha ^{'}},
\;\;\;M^2(\Xi )=\frac{3}{2\alpha ^{'}},
\eeq
\beq
M^2(\Delta )=\frac{5}{4\alpha ^{'}},\;\;\;M^2(\Sigma ^\ast )=\frac{13}{8
\alpha ^{'}},\;\;\;M^2(\Xi ^\ast )=\frac{2}{\alpha ^{'}},\;\;\;M^2(\Omega )=
\frac{19}{8\alpha ^{'}},
\eeq
(in (17) $M^2(\Sigma ^{'})\equiv (M^2(\Lambda )+M^2(\Sigma ))/2),$ and
in \cite{glue} for glueballs,
\beq
M^2(0^{++})=\frac{9}{4\alpha ^{'}},\;\;\;M^2(2^{++})=\frac{9}{2\alpha ^{'}},
\eeq
from collinear Regge trajectories for ordinary hadrons and glueballs, and the
cubic mass spectra associated with them. Eqs. (16)-(18) are in excellent 
agreement with experiment, and Eq. (19) with lattice QCD simulations, for 
$\alpha ^{'}\approx 0.85$ GeV$^{-2}.$

\section*{Concluding remarks}
We have shown that Regge phenomenology is consistent with the only scaling 
behavior for all hadron masses which is given in Eq. (7), which implies that
at finite temperature hadrons keep populating collinear trajectories the slope
of which depends on temperature but the intercepts do not. Some 
phenomenological models further specify the expression of the scaling law (7) 
in terms of the temperature-dependent gluon condensate, as in (15). The answer 
to the question about whether such a universal scaling is realized in the real
world, as well as its relation to the temperature-dependent quark condensate 
(as, e.g., in (1)), requires further theoretical and experimental study.

\section*{Acknowledgements}
The author wishes to thank T. Goldman and L.P. Horwitz for very valuable
discussions during the preparation of the present work.

\end{document}